\documentclass[]{raa}
\usepackage{graphicx,times}
\usepackage{natbib}
\usepackage{indentfirst}
\usepackage{ifpdf}

\providecommand{\bm}[1]{\mbox{\boldmath$#1$\unboldmath}}

\begin{document}
\title{A Time Series of Filament Eruptions Observed by Three Eyes from Space: From
   Failed to Successful Eruptions}
\volnopage{ {\bf 20XX} Vol.\ {\bf X} No. {\bf XX}, 000--000}
   \setcounter{page}{1}
   \author{Yuandeng Shen, \inst{1,2}
   Yu Liu \inst{1} and
   Rui Liu \inst{3}}
   \institute{National Astronomical Observatories/Yunnan Observatory, Chinese Academy of Sciences,Kunming 650011, China; {\it ydshen@ynao.ac.cn}\\
        \and
             Graduate University of Chinese Academy of Sciences, Beijing 100049\\
        \and
             Space Weather Research Laboratory, New Jersey Institute of Technology, University Heights, Newark, NJ 07102-1982, USA\\
\vs \no
   {\small Received [2010] [September] [21]; accepted [2010] [December] [21] }}

\abstract{ We present stereoscopic observations of six sequent eruptions of a filament in the active region NOAA 11045 on 2010 Feb 8, with the advantage of the {\sl STEREO} twin viewpoints in combination with the earth viewpoint from {\sl SOHO} instruments and ground-based telescopes. The last one of the six eruptions is with a coronal mass ejection, while the others are not. The flare in this successful one is more intensive than in the others. Moreover, the filament material velocity of the successful one is also the largest among them. Interestingly, all the filament velocities are found proportional to their flare powers. We calculate magnetic field intensity at low altitude, the decay indexes of the external field above the filament, and the asymmetry properties of the overlying fields before and after the failed eruptions and find little difference between them, indicating the same coronal confinement for the failed and the successful eruptions. The results suggest that, besides the confinement of coronal magnetic field, the energy released in low corona should be another crucial element for production of a failed or successful filament eruption. That is, only a coronal mass ejection can be launched away if the energy released exceeds some critical value, given the same coronal conditions.
\keywords{sun: filaments--- sun: corona ---sun: magnetic fields --- sun: flares --- sun: coronal mass ejections (CMEs)} }
   \authorrunning{Y. Shen et al.}
   \titlerunning{Failed Filament Eruptions Observed by Three Eyes from Space}
   \maketitle

\section{Introduction}
Filament activities have been extensively studied in solar physics. It is widely accepted that the magnetic field surrounding the filaments plays a key role in their formation, structure, and stability \citep{tang87,tand95,mart98}, and the filament eruptions are intimately associated with solar flares and CMEs \citep{lin00,zhan01,zhang02,huds06,yang08,shen10}. The three phenomena are suggested as different manifestations of a single physical process of coronal magnetic field \citep{harr95,prie02,lin03,jain10}, but the relationships among them are not well understood.

The failed filament eruptions have been documented in several studies (e.g. \citealt{ji03,alex06,gree07,liuy09}). Theoretical and statistical studies indicate that both the gradient of the overlying magnetic field with height and the field intensity at low altitude are key factors in determining the flux rope's ultimate fate \citep{toro05,klie06,fan07,liuy08}. According to these studies, slow magnetic field decreasing gradient and strong field intensity at low altitude are in favor of failed eruptions by strong confinement of the coronal magnetic field. Recently, \cite{liuy09} investigate a failed filament eruption case in which two producing filaments are covered by asymmetrical coronal loops formed due to imbalanced magnetic flux; one of the filaments erupts at first and then interacts with the other \citep{liuy10}; they merge together and move out resembling an EUV jet but fail to escape from the sun. Based on these observational facts, they calculate and compare the confinement ability of symmetric and asymmetric fields and find that the magnetic confinement of an asymmetric field is stronger than that of a symmetric one. Therefore, they suggest that an asymmetric background field is an impact factor leading to failed filament eruptions. The above studies imply that coronal magnetic field distribution and configuration are important for us to understand the process of failed filament eruption. However, for the relation between failed and successful filament eruptions in a same coronal condition, no observational results have been presented for it.

In this paper, we study a filament and its six eruptions including both failed and successful cases. They are associated with a series of B- and C- class {\sl GOES} soft X-ray (SXR) flares occurred in NOAA AR 11045 (N22, W00) on 2010 Feb 8. The filament is observed struggling to escape the solar surface, and it fails a few times due to the obvious confinement of the overlying magnetic fields, but it succeeds finally when a C6.2 flare takes place. This successful eruption had led to an earth-directed gradual CME which caused some mild geomagnetic activity about 3.5 days later. We analyze multi-angle and multi-wavelength data for these events in order to fully understand what are the main factors for a filament eruption to become failed or successful. The paper is arranged as follows: In Section 2, we describe the data used in this paper, Section 3 contains the results of this study, and discussion and conclusions are presented in Section 4.

\section{Observations}
The observations used in this work are listed as follows:

1. Full-disk H$\alpha$ images from the H$\alpha$ telescope of Yunnan Astronomical Observatory (YNAO). We use the H$\alpha$ line-center images with a cadence of 1 minute and a pixel size of 1$^{''}$, recorded by a 3k $\times$ 2k 16 bit CCD camera. The YNAO's observation started at 01:21 UT and ended at 08:49 UT on 2010 Feb 8.

2. Full-disk line-of-sight magnetograms and EUV 195 \AA\ images from the Michelson Doppler Imager (MDI) \citep{sche95} and the Extreme Ultraviolet Telescope (EIT) \citep{dela95} aboard the {\sl Solar and Heliospheric Observatory (SOHO)}. The MDI magnetograms with a pixel size of 2$^{''}$ and a 96 minutes cadence, while the EIT 195 \AA\ images with a pixel size of 2.6$^{''}$ and the cadence is 12 minutes.

3. Full-disk EUV 195 and 304 \AA\ images are taken by the Extreme Ultraviolet Imager (EUVI) \citep{Wue04} of the Sun Earth Connection Coronal and Heliospheric Investigation (SECCHI) \citep{How08} on board the {\sl Solar Terrestrial Relations Observatory} ({\sl STEREO}) spacecraft \citep{Kai08}. On Feb 8, {\sl STEREO} Ahead ({\sl STEREO-A}) and Behind ({\sl STEREO-B}) spacecrafts provided continuous images simultaneously from two different viewpoints with which the separation angle was about $135^{\circ}$. The images with a pixel resolution of $1.6^{''}$ and a time cadence of 5 (10) minutes for 195 (304) \AA. The coronagraph total brightness images, which were taken by the inner coronagraphs (COR1) \citep{thom03} of SECCHI aboard the {\sl STEREO}, with a field of view (FOV) range from 1.4 to 4 solar radii; the pixel resolution is 15$^{''}$ and the cadence is 5 minutes. In the following, we will refer to the images observed by EUVI (COR1) of SECCHI on board the {\sl STEREO-A} and {\sl STEREO-B} as EUVI-A (COR1-A) and EUVI-B (COR1-B) for convenience.

4. The {\sl Reuven Ramaty High Energy Solar Spectroscopic Imager (RHESSI)} \citep{linr02} images are reconstructed by use the Pixon algorithm \citep{metc96} of the {\sl RHESSI} data analysis software in SolarSoftWare (SSW). The Pixon algorithm provides a significant better photometry, accurate position, and allows for the detection of fainter sources. In the image reconstruction, detectors 3-8 are used and the integration time is 1 minute around the {\sl RHESSI} flare peak.

5. {\sl Transition Region and Coronal Explored (TRACE)} white light image \citep{Han99}. In the study, only one {\sl TRACE} white light image is available and used.

\section{Analysis and Results}
\subsection{The Failed Filament Eruptions}
\begin{table}[b!!!]
\small
\centering
\begin{minipage}[]{95mm}
\caption[]{ {\bf Information of the six {\sl GOES} SXR Flares Accompanying the Filament Eruptions in AR 11045 on 2010 Feb 8}}
\label{Table 1}
\end{minipage}
\tabcolsep 2.5mm
 \begin{tabular}{cccccccc}
  \hline\noalign{\smallskip}
NO. &  Start (UT) & Peak (UT) & End (UT) & Class & Location (from Earth)    \\
  \hline\noalign{\smallskip}
1 &  00:10 & 00:16 & 00:18 & C1.4 & N22E02  \\ 
2 &  01:10 & 01:14 & 01:16 & B5.8 & N21E01  \\
3 &  01:23 & 01:29 & 01:31 & B7.0 & N21E01  \\
4 &  02:22 & 02:37 & 02:40 & B7.4 & N22W01  \\
5 &  02:46 & 02:52 & 02:56 & B8.3 & N22W00  \\
6 &  03:08 & 03:17 & 03:23 & C6.2 & N23W00  \\
  \noalign{\smallskip}\hline
\end{tabular}
\end{table}

\begin{figure}[b!!!]
\centering
\includegraphics[width=0.75\textwidth]{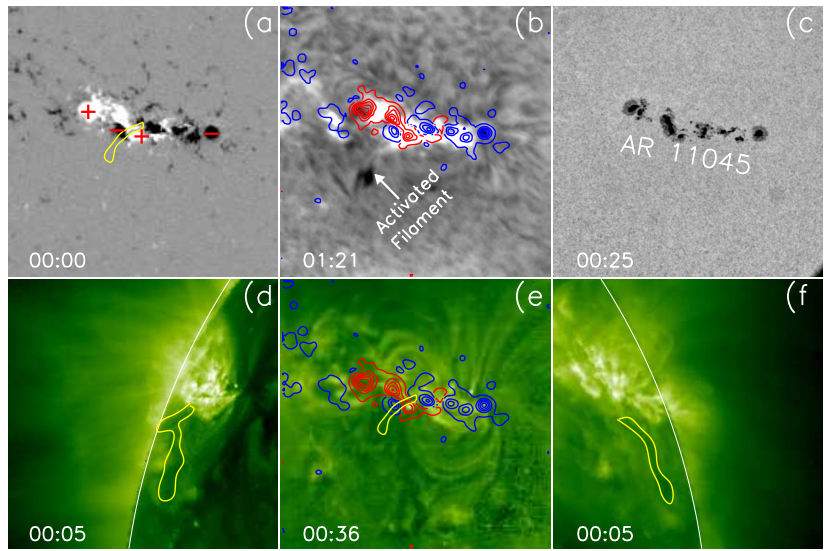}
\begin{minipage}[]{110mm}
\caption{ MDI line-of-sight magnetogram (a), YNAO H$\alpha$ line-center (b), {\sl TRACE} white-light (c), and EUVI-A (d), EIT (e), and EUVI-B (f) 195 \AA\ images before the filament eruptions. The active region is marked out in the {\sl TRACE} white-light image. Contours of MDI magnetic fields are overlaid on the H$\alpha$ and EIT images, with which positive (negative) fields are in red (blue). The white arrow indicates the activated filament, meanwhile, the filament in EUV 195 \AA\ images are outlined as yellow contours. In the frame (a), yellow outline is the filament outline in frame (e), and the red + (-) signs label the positive (negative) magnetic polarities. The white curves mark the disk limb (the same in the subsequent figures), and the FOV is $380^{''} \times 380^{''}$ for each frame.}
\end{minipage}
\label{Fig1}
\end{figure}

On 2010 Feb 8, the active filament was located at the center of NOAA AR 11045 (N22, W00), a rapidly evolving active region with emerging magnetic flux around its core. During 00:00 to 04:00 UT, the filament underwent six eruptions, including five failed and one successful, and each was accompanied by a small flare. All the associated flares were in short duration and were of B- or C- class, and they took place nearly at the same place (cf. Table 1). Fig. 1 is an overview of the active region in multi-angle and multi-wavelength before the eruptions. The active region exhibited a quadrupole configuration in the MDI magnetogram while the opposite polarities interlaced with each other (see Fig. 1a). The filament is discernible both in the H$\alpha$ line-center and the EUV 195 \AA\ images (see the white arrow in the frame (b) and yellow contours in frames (d)--(f)). The contours of MDI line-of-sight magnetic field in Fig. 1 indicate that the filament is located on the magnetic neutral line but is slightly over the negative polarity region; such magnetic field configuration is thought to be in favor of some filament instability \citep{mart98}. Owing to the lack of H$\alpha$ data before 01:21 UT on Feb 8, the filament in H$\alpha$ line-center wavelength at 01:21 UT has already been activated and shown as a thick and dark feature in the H$\alpha$ image. On the other hand, the filament exhibited different shapes in different aspect angles in the same wavelength images, which was mainly due to the projection effect from different viewpoints (see the yellow contours in frames (d)--(f) in Fig. 1).

\begin{figure}[b!!!]
\centering
\includegraphics[width=0.75\textwidth]{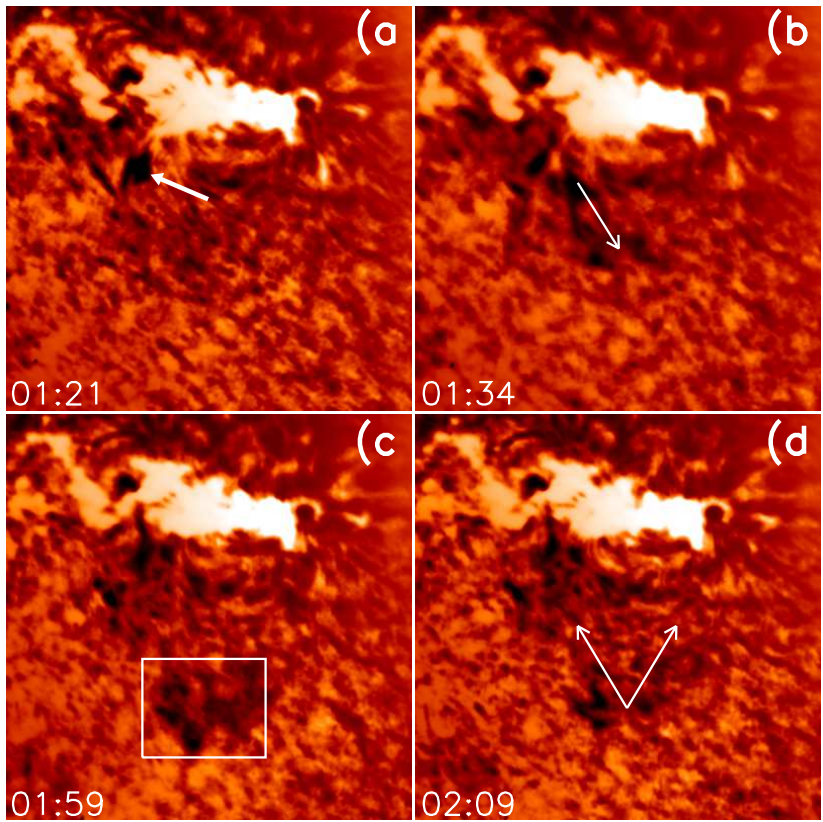}
\begin{minipage}[]{110mm}
\caption{ YNAO H$\alpha$ line-center images show the filament's main evolving process. The thick white arrow in the frame (a) indicates the activated filament before eruption, white thin arrow in the frame (b) shows the filament's erupting direction, and the two white thin arrows in (d) show the draining direction of the accumulated plasmoids. The white rectangle is the region where the H$\alpha$ light curve is measured and displayed in Fig. 3. The FOV is $400^{''} \times 400^{''}$ for each frame.}
\end{minipage}
\label{Fig2}
\end{figure}

The first two eruptions are untraceable in H$\alpha$ due to no H$\alpha$ data available. Fortunately, YNAO made good observations on the rest of the events. In fact, the preceding five filament eruptions could be identified as failed eruption events in combination with other observations. Here the third one is selected as an example to illustrate the failed eruption process because it is the best observed one of the six events (cf. Fig. 2). The filament showed as a dark and thick feature at 01:21 UT just prior to the associated flare start (01:23 UT) (see the thick white arrow in Fig. 2a), which might indicate the enhanced helical motions in the filament body \citep{Mar80, Sch91}. By 01:34 UT, the erupting filament moved out in a configuration similar to a chromospheric surge or a coronal jet along a loop-shaped trajectory (see Fig. 2b). Such filament eruptive behavior suggests that the filament was asymmetrically confined by the overlying loops, and its initial moving direction was along the lower part of the loops \citep{liuy09}. It is interesting to find that the moving plasmoids ceased to move out at a place about 185 Mm away from the primary location. On the contrary, they piled up there (see Fig. 2c) and lasted for a long time before draining back to the solar surface along both legs of the overlying loops (see two white arrows in Fig. 2d). The light curve of the mass piling region indicates that the intensity drops suddenly to about 65\% of the initial level and then goes through a long gradual recovery process (about 50 minutes, see Fig. 3). Obviously, the sudden drop of the intensity is caused by the intrusion of the dark cold mass of the erupting filament, and the slow drainage motion of the accumulated plasmoids brings about the gradual recovery process.

\begin{figure}[t!!!]
\centering
\includegraphics[width=0.75\textwidth]{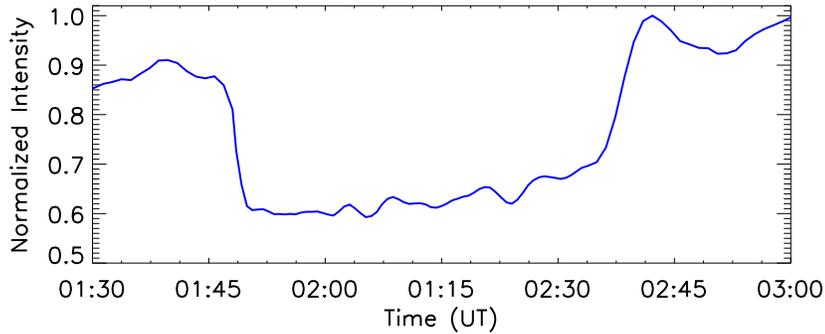}
\begin{minipage}[]{110mm}
\caption{Normalized light curve at the location where the ejecting filament mass was piling up seen from H$\alpha$ observations.}
\end{minipage}
\label{Fig3}
\end{figure}

\begin{figure}[t!!!]
\centering
\includegraphics[width=0.75\textwidth]{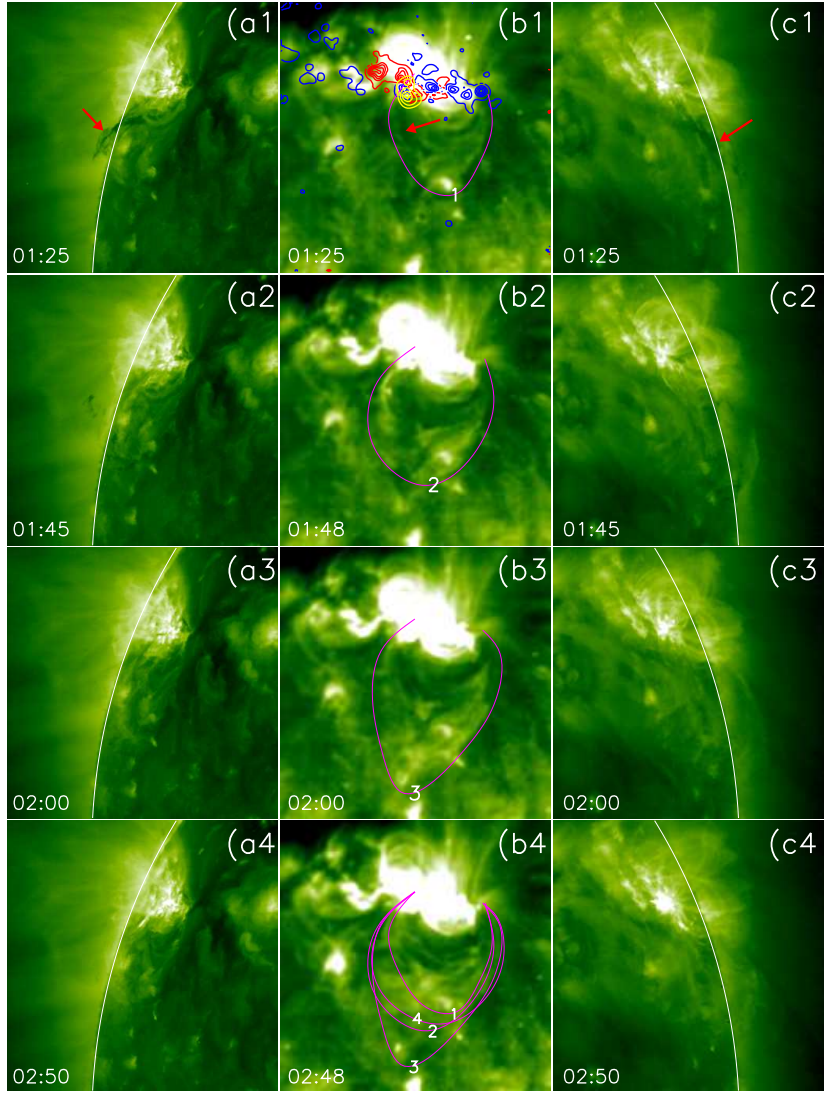}
\begin{minipage}[]{110mm}
\caption{Time sequences of EUVI-A (a1-a4), EIT (b1-b4), and EUVI-B (c1-c4) 195 \AA\ images. The red arrows indicate the dark filament and the fuchsia curves represent the profile of the overlying loops. The MDI magnetic field at 01:39 UT is overlaid on frame (b1) as red (positive) and blue (negative) contours, while the yellow contours are the {\sl RHESSI} HXR sources at 01:27 UT, and the contour levels are 35\%, 50\%, 70\%, and 90\% of the maximum brightness. The FOV is $450^{''} \times 450^{''}$ for each frame.}
\end{minipage}
\label{Fig4}
\end{figure}

Fig.4 shows the failed filament eruption process in time sequence of EUVI-A, EIT, and EUVI-B 195 \AA\ raw images. The outer profile of the overlying loops are plotted as fuchsia curves on the EIT 195 \AA\ images. All curves at different time are overlaid on the 02:48 UT frame for comparison and order numbers are also labeled. The {\sl RHESSI} HXR sources show the associated flare was occurring near the loop's east footpoint. In fact, this region is the magnetic flux emerging region where magnetic cancellations take place.

The most impressive characteristics in EUV 195 \AA\ are the dynamic behavior of the filament eruption coupling with the overlying loops. The cool, dense filament exhibited as dark, absorption structure in EUV wavelength for its relatively low temperature nature (see the red arrows in Fig. 4). By 01:25 UT, the filament has already been in the eruption process. It was moving along but below the overlying loops instead of escaping (see the red arrow in Fig. 4b1); it manifests that the erupting filament was confined by the overlying loops. Similar to the H$\alpha$ line-center observations, the moving filament material stopped and accumulated near the loop top. In the course of the filament eruption, the overlaying loop system experienced an expansion and contraction process, which should be due to the interaction of the erupting filament (compare curves in Fig. 4b4). At 02:00 UT, the stretched loops exhibited as a cusp structure in EIT 195 \AA\ images, and the top section was obviously brighter than its both legs (see Fig. 4b3). The brightening is thought to be due to the accumulating hot plasmoids. After this moment, the loops begin to contract inward slowly. It is worthwhile to note that the loop expansion (contraction) process observed in EUV 195 \AA\ is in well correspondence to the filament outward (draining back) motion observed in H$\alpha$ line-center on temporal and spatial scales; it implies that the filament eruption is indeed failed at different levels of the solar atmosphere. The same evolving process can be identified from both EUVI-A and EUVI-B 195 \AA\ images, but the projection effect need to be added in (see the left and the right row of Fig. 4).

We can not find any correspondent CME for this event from the databases of COR1-A, COR1-B, and LASCO \citep{brue95}; these coronagraphs are monitoring CMEs continuously from three different angles. Furthermore, on 2010 Feb 8 the active region was located near the solar central meridian passage in {\sl SOHO} observation, while it was near the disk limb when observed from the {\sl STEREO} twin viewpoints. Therefore, if a CME originates from this active region, it is expected to be recorded at least by one coronagraph. Based on the above analysis, we conclude that this filament eruption is a failed event definitely.

\begin{figure}[t!!!]
\centering
\includegraphics[width=0.75\textwidth]{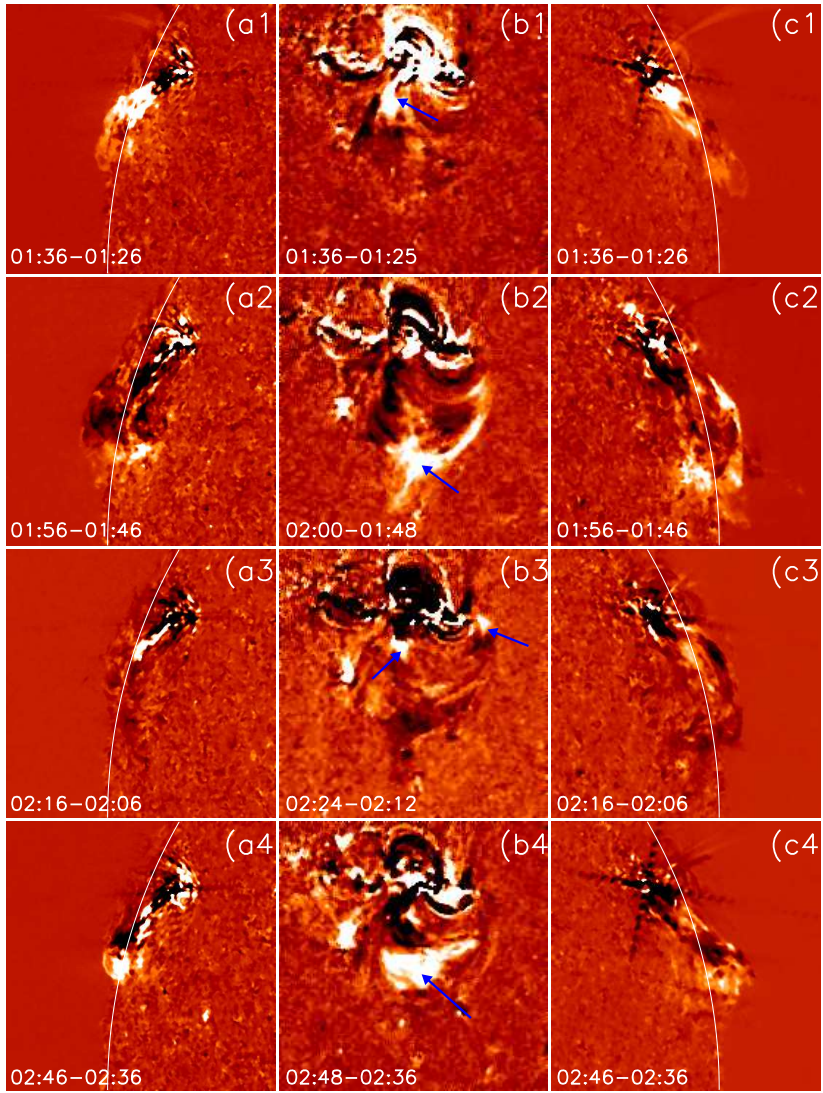}
\begin{minipage}[]{110mm}
\caption{Time sequences of running difference images of EUVI-A 304 \AA\ (a1-a4), EIT 195 \AA\ (b1-b4), and EUVI-B 304 \AA\ (c1-c4). Blue arrows indicate the moving erupting filament, and the FOV is $500^{''} \times 500^{''}$ for each frame.}
\end{minipage}
\label{Fig5}
\end{figure}

\begin{figure}[b!!!]
\centering
\includegraphics[width=0.75\textwidth]{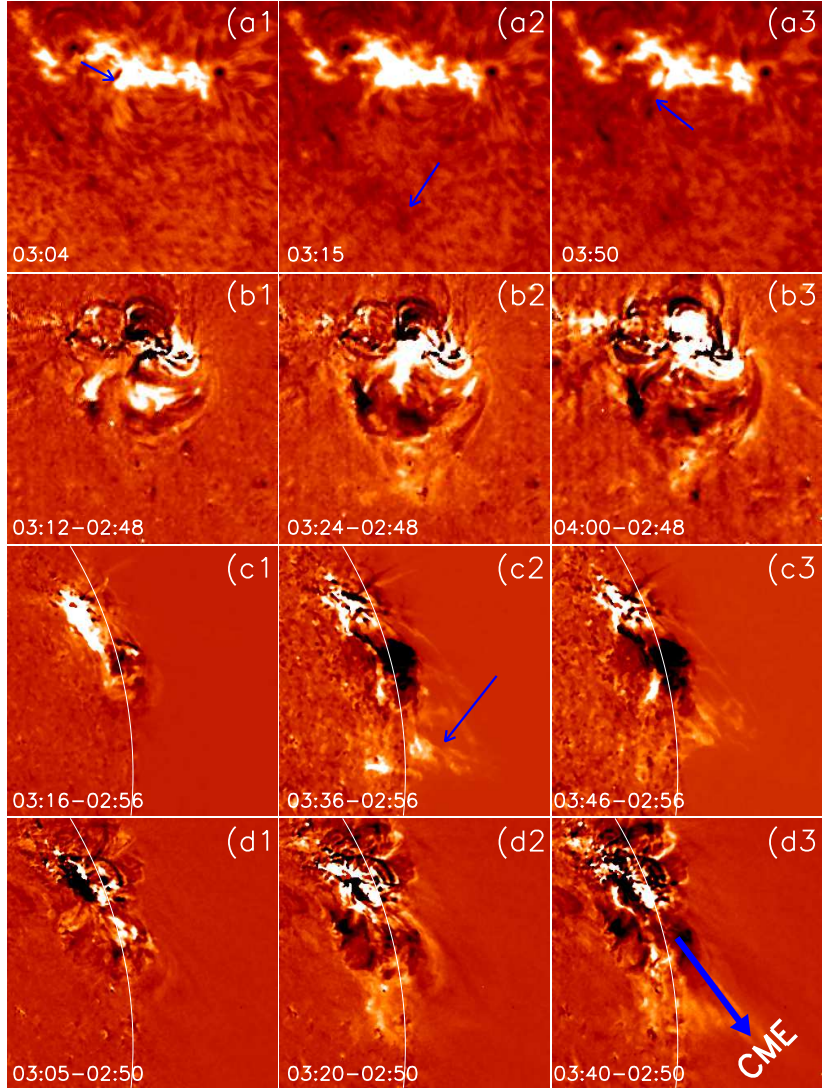}
\begin{minipage}[]{110mm}
\caption{YNAO H$\alpha$ line-center images(a1-a3), and the fixed-base difference images of EIT 195 \AA\ (b1-b3), EUVI-B (c1-c3) 304, and 195 \AA\ (d1-d3), respectively. The short blue arrows indicate the filament at different time in H$\alpha$ wavelength. The long thin blue arrow indicates the broken site of the filament in EUVI-B 304 \AA\ while the long thick blue arrow show the filament escaping part in EUVI-B 195 \AA\ images. The FOV is $600^{''} \times 600^{''}$ for all frames.}
\end{minipage}
\label{Fig6}
\end{figure}

The EUVI-A and EUVI-B 304 \AA\ and EIT 195 \AA\ running difference images are shown in Fig. 5, demonstrating the kinematics of the overlying loops confining the filament. We trace the white features, which represent the moving filament material, in time sequences images. It is clear to see that the filament started to erupt from its primary site and kept moving along the overlaying loops. However, it stopped when it reached the loop top. The top and the west parts of the loops were obviously brightened because of the injection of the heated filament mass around 02:00 UT (see the blur arrows in top two rows of Fig. 5). The blue arrows, in the frame (b3) of Fig. 5, point to two white patches where the receding filament material was falling to. Meanwhile, the dimming region, formed at the loop top section, indicates mass evacuated away there (see Fig. 4c3); It is mainly due to the downward moving of the accumulated plasmoids whin the confined space of the overlying loops. Later, a more brightening patch appeared at 02:48 UT near the loop top (see the blue arrow in Fig. 5b4); it was caused by another contiguous filament eruption. Obviously, the brightening patch was now closer to the original location of the filament compared to Fig. 5b2. This indicates that the whole loop system has contracted after expanding during the previous eruption event (also refer to the fuchsia curves in Fig. 4b4), therefore, the erupting filament was confined at a lower height and endured more confinement of the contracted loops because the magnetic fields around the loop top should become more horizontal \citep{huds08,wang10}.

Based on the above analysis of the eruption events (No.3 and 4 of Table 1), using multi-angle and multi-wavelength observations we believe that the eruptions are really failed and have been well confined by closed coronal fields. It should be pointed out that studies using only single viewpoint data may result to incorrect conclusion. The occulter of a coronagraph may ward off a low-brightness CME, especially when the eruption occurs from the disk center and heads along the line-of-sight direction to the coronagraph. In the following subsection, by using the subsequent successful filament eruption event, we will show how important to use stereoscopic data for the more reliable CME detecting.

\begin{figure}[b!!!]
\centering
\includegraphics[width=0.75\textwidth]{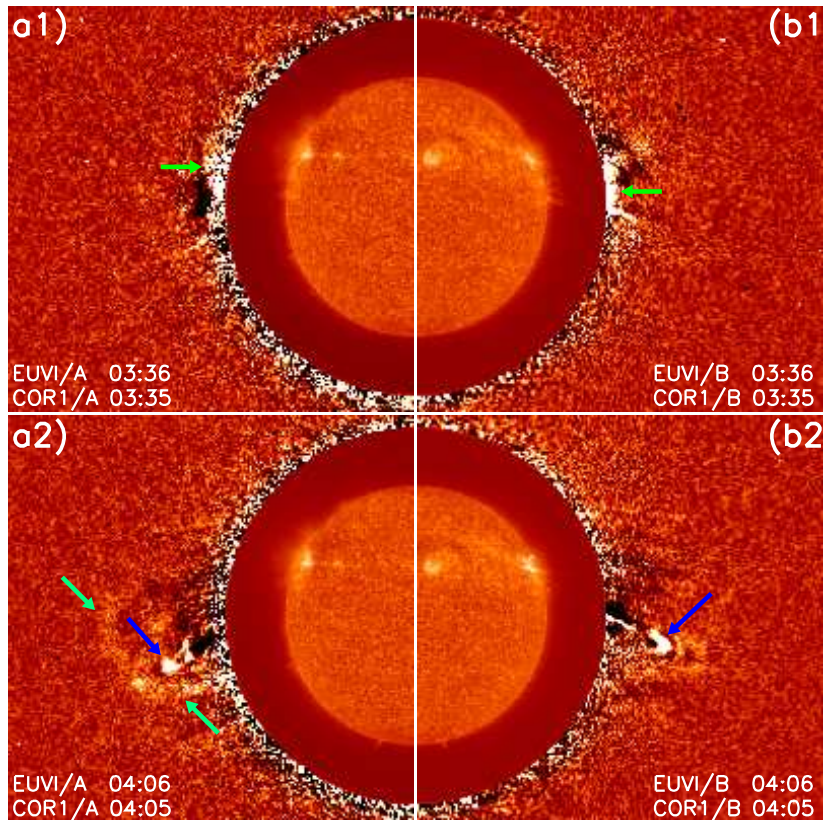}
\begin{minipage}[]{110mm}
\caption{The composite images of inner EUVI 304 \AA\ and outer COR1 difference images, showing the associated CME of the partial filament eruption. The left row is {\sl STEREO-A} data and the right is {\sl STEREO-B}.}
\end{minipage}
\label{Fig7}
\end{figure}

\subsection{The Successful Filament Eruption}
After a series of failed eruptions listed in Table 1 (No.1--5), a subsequent successful filament eruption was observed by {\sl STEREO} EUVI and COR1 telescopes (No.6, Table 1). The eruption occurs following the end of the fifth failed eruption and was accompanied by a {\sl GOES} SXR C6.2 flare. Unlike the failed eruptions analyzed in the above sections, part of the filament mass successfully escapes from the sun and a CME was detected by COR1-A and COR1-B. The event process is clearly shown in Fig. 6. The top row shows the filament before, during, and after the eruption in H$\alpha$ images. It starts to erupt around 03:04 UT and moves out along a loop-like trajectory in a way similar to the preceding failed eruptions (see the short blue arrow in Fig. 6a2). The filament, however, reformed near its primary location shortly after the  eruption (see the short blue arrows in Fig. 6a3), which could be regards as a direct evidence for partial filament eruption \citep{trip09}. In EUV observation, the evolving process of the filament in EIT 195 \AA\ is also similar to the preceding failed eruptions. However, the upper part of the filament is identified to be ejected into interplanetary space in the EUVI-B 304 \AA\ difference images, and the broken site is indicated by the long thin blue arrow in Fig. 6c2. The bottom row is the EUVI-B 195 \AA\ difference images, from which the opening process of the overlying loops could be identified and the escaping part is indicated by the thick long blue arrow (Fig. 5d3). Although some surface phenomena such as EIT wave and corona dimming are observed associated with this event from {\sl SOHO}/LASCO database, no CME can be found from it.

Following the partial eruption, however, an associated CME was obviously observed from side by COR1-A and COR1-B although it was not catched by the LASCO telescope. Fig. 7 shows the CME evidence in the composite images. A close temporal and spatial relationship between the CME and the erupting filament can be identified from the images. The bright front of the CME first appeared at 03:36 UT in both coronagraphs' FOV (see the two thick green arrows in Fig. 7). The typical three components (viz., a bright front, a dark cavity and a bright compact core) of the CME are observed clearly in COR1-A at 04:05 but slightly fainter in COR1-B for the CME front. The bright front (compact core) is indicated by the green (blue) arrows in the bottom row of Fig. 7. The CME average velocity measured from COR1-A is about 548km/s, i.e., a gradual CME.

Because it is always easier to observe a CME initiated at the disk limb than on the disk, it is not surprised that the CME event can not be detected by the LASCO telescope. The case presented in this section is a good example to show the importance of using stereoscopic observations. If only using the data from the EIT and LASCO instruments, one might mistakenly consider it as a failed filament eruption.

\begin{figure}[b!!!]
\centering
\includegraphics[width=0.75\textwidth]{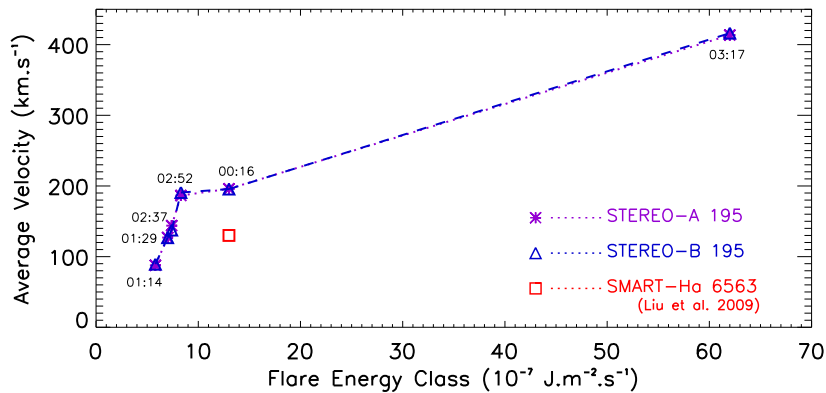}
\begin{minipage}[]{110mm}
\caption{ Average velocity of filament material as a function of {\sl GOES} SXR flare energy class. The asterisks (triangles) represent {\sl STEREO-A} ({\sl STEREO-B}) data. The single isolated red square is the data taken from \cite{liuy09}. The peak time of the flares are also denoted.}
\end{minipage}
\label{Fig8}
\end{figure}

\subsection{Relation between Flare Power and Filament Ejection}
Fig. 8 shows the filament average velocity during the rising phases as a function of the SXR flare energy class. The data is measured from EUVI-A and EUVI-B 195 \AA\ images respectively. The data distribution of the C1.3 flare, associated with a failed filament eruption studied by \cite{liuy09}, deviates from other points obviously, which is possibly due to the fact that its velocity was measured from H$\alpha$ wavelengths. The filament average velocity of the C1.3 flare event does not exceed the calculated values for the correspondent class of flares in our study. Generally, a positive correlation is found between filament velocity and flare power based on the current data. From the result we can conclude that a filament tends to obtain more kinetic energy probably transformed from the magnetic energy released in a bigger flare associated by reconnection than that from a smaller one. This results support the statistical study by \cite{mori10}, who find that the thermal energy density of the overlying arcade increases with the filament's mechanical energy density. Moreover, if the efficiency of the filament mass transfer is high enough (represented by high velocity) to destroy the original mechanical equilibrium of the flux rope system in the corona, then a reconstructing process of the coronal magnetic field is expected to occur in order to help the system to reach another equilibrium, by releasing free magnetic flux helicity in the form of a CME.

\begin{figure}[b!!!]
\centering
\includegraphics[width=0.75\textwidth]{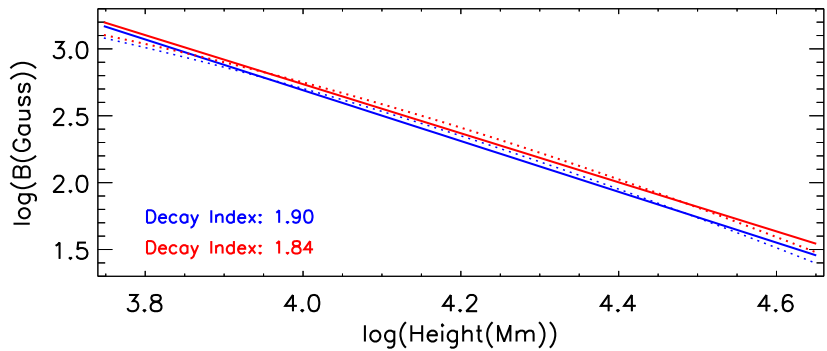}
\begin{minipage}[]{110mm}
\caption{ A fitting of the calculated magnetic field strength and height measured from the photosphere. Here the strength and height are in log units. The blue (red) dotted line is horizontal component of the potential field before (after) the failed eruptions, while the solid lines are the result of a linear fitting to the data.}
\end{minipage}
\label{Fig9}
\end{figure}

\subsection{External Fields Over The Erupting Filament}
The gradient of the external fields with height is believed to be an important factor for diagnosis of failed or successful filament eruptions \citep{klie06,liuy08,liuc10}. It is usually expressed as the so-called decay index, defined as n = {\sl - d}ln({\sl B})/{\sl d}ln({\sl h}). Here {\sl B} is the poloidal component of the external potential field above the filament and {\sl h} is the height measured from the photosphere. The potential field model is used to approximate the external magnetic field, and the poloidal component of the external potential field is represented by the horizontal components of the extrapolated potential field. The potential fields at 00:00 UT (before the failed eruptons) and 03:15 UT (after the failed eruptions) are calculated using the NLFFF (A Non-Linear Force-Free-Field Extrapolation Program) code \citep{whea00}, in which the potential field calculations are made in Cartesian geometry by a Green's function approach \citep{chiu77}. The decay indexes, which are derived from a linear fitting to the data, before and after the failed eruptions are 1.90 and 1.84 (see Fig. 9). In our calculation, the height range is set to 42 to 105 Mm (height range for filament instability \citep{liuy08}) above the photosphere over the filament eruption place. We also calculate the average decay index in the same height range by setting the step size as the grid spacing ($\sim$ 1.41 Mm) in the potential field extrapolation and the results are 1.95 and 1.89. The decay index differences between the two moments is very small. Whether a filament eruption is failed or successful, there are two other important impact factors: the strength of horizontal field at low altitude and the asymmetric confinements of the overlying fields \citep{liuy08,liuy09}. Hence we compare the horizontal field at a height of 42 Mm, which is the average height of a eruptive filament, at the two moments based on the potential calculation. The values are 21.5 and 22 Gauss respectively. They are much the same for before and after the failed eruptions. Although the filament was asymmetrically confined by the overlying loops \citep{liuy09}, it is not the primary influencing factor for the present study because no obvious morphological change between the two special moments could be identified from the observations. In addition, the ratios of the horizontal field strength at the filament position to the symmetric position at the two moments also show few difference. The above results of calculation show that the failed eruptions involving the associated flares do not affect the confinement ability of the external fields significantly. Hence all the eruptions endured almost the same coronal confinement of the external fields. In addition, the indexes are very close to the upper end of theoretical critical decay index (1.5-2.0) for successful eruptions \citep{klie06}. It implies that the magnetic system involving the eruptive filament is on the edge of instability. Since the flares in the failed eruptions are not powerful enough, these results seem to suggest that more energy is needed to cause a successful eruption. The successful eruption associated with a relative more powerful C6.2 flare in the present study is a good case in the point.

\section{Conclusions and Discussion}
The filament eruptions studied in the present paper are associated with a series of B- and C-class flares, in which five are identified as failed eruptions with no CME association and one is a successful eruption with clear evidence of a CME's occurrence. The eruptions are possibly attributed to the emerging magnetic flux around the active region core where magnetic cancellations are observed between the pre-existing and the newly emerging flux. The processes of the eruptions are well observed synchronously by ground-based and space-based telescopes, from which multi-angle and multi-wavelength data are available. It should be noted that the successful eruption is a partial filament eruption, and its associated CME is not detected by {\sl SOHO}/LASCO from the earth direction but by two other directions from space by {\sl STEREO}/COR1 A and B instruments. During the failed eruption period, the filament is found to be confined by an overlying loop system. The neutral line of the active region, where the filament is located, is very close to the east footpoint of the overlying coronal loops. In the filament eruptions, the mass is observed moving along the overlaying loops from the east root in a shape resembling an EUV jet or an H$\alpha$ surge. The filament is observed struggling to escape during the failed eruptions. And the tops of the overlying coronal loops are found to be dragged forward and brightened. The filament eruption finally succeeds to escape the sun when a powerful C6.2 flare takes place, which is accompanied by a CME launched away.

We find that the velocity profile of the erupting filament has a good proportional relation with the energy class of the associated {\sl GOES} SXR flares. Because the successful filament eruption is associated with a most intensive flare, it seems that the flare power is an important index in the physical process causing a failed or successful filament eruption. Unfortunately, due to the limited event samples, it is difficult to get the exact critical value for the flare power, above which a successful filament eruption can be expected for this active region. But we believe that a flare with high energy power is important for us to forecast a successful filament eruption. Based on the potential field extrapolation, we compare the horizontal field at low altitude before and after the failed eruptions and find that the field strengthes are much the same. On the other hand, the asymmetric confinement of the overlying fields show unconspicuous change at the two moments considered. These results together provide the indication of the same coronal condition for the failed and the successful eruptions.

The failed filament eruptions in our observations are probably due to the following reasons: (1) the stronger field intensity at low altitude, (2) low magnetic field gradient (decay index) of the overlying loops with height, (3) asymmetrical magnetic confinement of the overlying fields, and (4) high enough kinetic energy for the erupting filament mass. Since the magnetic field intensity, decay index and asymmetry properties are almost the same during the eruptions, the last reason should be a new factor found in this study because the successful eruption event is associated with the most powerful flare, which can supply more kinetic energy to the filament mass by strong outflow from the reconnection involved.

Moreover, some failed filament eruptions have been reported recently in several studies (e.g., \citealt{ji03,liuy09}), but all of them are based on single aspect angle observations. When eruptions occurred, the associated CMEs might be warded off by coronagraph's occulter, so that it may not appear in the coronagraph's FOV for the case of a faint CME launched. To overcome this drawback, multi-angle and multi-wavelength observations are necessary. The eruption in our study well speaks volume for this issue. Hence we should maintain cautions on the single angle observations not only for identifying failed filament eruptions but also for analyzing other various forms of solar events.

\normalem
\begin{acknowledgements}
We thank the YNAO, {\sl RHESSI}, {\sl TRACE}, {\sl GOES}, {\sl STEREO} SECCHI, and the {\sl SOHO} MDI, EIT and LASCO teams for data support, and the referee for his/her constructive comments. This work is supported by the Chinese foundations MOST (2011CB811400) and NSFC (10933003, 11078004, and 11073050).
\end{acknowledgements}

\label{lastpage}

\begin{thebibliography}{99}
\small \setlength{\itemindent}{-3mm} \setlength{\itemsep}{-0.5mm}
\setlength{\baselineskip}{4.5mm}
\bibitem[{Alexander}{~et~al.}(2006)]{alex06}                         Alexander, D., Liu, R., \& Gilbert, H. R. 2006, \apj, 635, 719
\bibitem[{Bruechner}{~et~al.}(1995)]{brue95}                         Bruechner, G. E., Howard, R. A., Koomen, M. J., et al. 1995, \solphys, 162, 357
\bibitem[{Chiu \& Hilton}(1977)]{chiu77}                             Chiu, Y. T., \& Hilton, H. H. 1977, \apj, 212, 873
\bibitem[{Delaboudini$\grave{\rm e}$re}{~et~al.}(1995)]{dela95}      Delaboudini$\grave{\rm e}$re, J.-P., Artzner, G. E., \& Brunaud, J. 1995,\solphys, 162, 357
\bibitem[{Fan \& Gibson}(2007)]{fan07}                               Fan, Y., \& Gibson, S. E. 2007, \apj, 9, 131
\bibitem[{Green \& Kliem}(2007)]{gree07}                             Green, L., M., kliem, B., T$\ddot{\rm o}$r$\ddot{\rm o}$k, T., van Driel-Gesztelyi, L., \& Attrill, G. D. R. 2007, \solphys, 246, 365
\bibitem[{Handy} {et~al.}(1999)]{Han99}                              Handy, B. N., Bookbinder, J., Deluca, E., et al. 1999, \solphys, 187, 229
\bibitem[{Harrison}(1995)]{harr95}                                   Harrison, R. A. 1995, \aap, 304, 585
\bibitem[{Howard} {et~al.}(2008)]{How08}                             Howard, R. A., Moses, A., Vourlidas, J. S., et al. 2008, \ssr, 136, 67
\bibitem[{Hudson}{~et~al.}(2006)]{huds06}                            Hudson, H. S., Bougeret, J.-L., \& Burkepile, J. 2006, \ssr, 123, 13
\bibitem[{Hudson}{~et~al.}(2008)]{huds08}                            Hudson, H. S., Fisher, G. H., \& Welsch, B. T. 2008, in ASP Conf. Ser. 383, Subsurface ang Atmospheric Influences on Solar Activity, ed. R. Howe et al. (San Francisco, CA: ASP), 221
\bibitem[{Jain}{~et~al.}(2010)]{jain10}                              Jain, R., Aggarwal, M., \& Kulkarni, P. 2010, \raa, 10, 473
\bibitem[{Ji}{~et~al.}(2003)]{ji03}                                  Ji, H. S., Wang, H. M., Schmahl, E. J., Moon, Y.-J., \& Jiang, Y. C. 2003, \apjl, 595, L135
\bibitem[{Kaiser} {et~al.}(2008)]{Kai08}                             Kaiser, M. L., Kucera, T. A., Davila, J. M., et al. 2008, \ssr, 136, 5
\bibitem[{Kliem \& T$\ddot{\rm o}$r$\ddot{\rm o}$k}(2006)]{klie06}   Kliem, B., \& T$\ddot{\rm o}$r$\ddot{\rm o}$k, T. 2006, \prl, 96, 255002
\bibitem[{Lin \& Forbes}(2000)]{lin00}                               Lin, J., \& Forbes, T. G. 2000, \jgr, 105, 2375
\bibitem[{Lin}{~et~al.}(2003)]{lin03}                                Lin, J., Soon, W., \& Baliunas, S. L. 2003, NewA Rev., 47, 53
\bibitem[{Lin}{~et~at.}(2002)]{linr02}                               Lin, R. P., Dennis, B. R., Hurford, G. J., et al. 2002, \solphys, 210, 3
\bibitem[{Liu}{~et~al.}(2010)]{liuc10}                               Liu, C., Lee, J., jing, J., Liu, R., Deng, N., \& Wang, H. M. 2010, \apjl, 721, L193
\bibitem[{Liu}(2008)]{liuy08}                                        Liu, Y. 2008, \apj, 679, L151
\bibitem[{Liu}{~et~al.}(2009)]{liuy09}                               Liu, Y., Su, J., Xu, Z., Lin, H., Shibata, K., \& Kurokawa, H. 2009, \apjl, 696, L70
\bibitem[{Liu}{~et~al.}(2010)]{liuy10}                               Liu, Y., Su, J. T., Shen, Y. D., \& Yang, L. H. 2010, in IAU Symp. 264, Solar and Stellar Variability: Impact on Earth and Planets, ed. A. G. Kosovichev, A. H. Andrei, \& J.-P. Rozelot, 99
\bibitem[{Martin}(1980)]{Mar80}                                      Martin, S. F. 1980, \solphys, 68, 217
\bibitem[{Martin}(1998)]{mart98}                                     Martin, S. F. 1998, \solphys, 182, 107
\bibitem[{Metcalf}{~et~al.}(1996)]{metc96}                           Metcalf, T. R., Hudson, H. S., Kosugi, T., Puetter, R. C., \& Pina, R. K. 1996, \apj, 466, 585
\bibitem[{Morimoto}{~et~al.}(2010)]{mori10}                          Morimoto, T., Kurokawa, H., Shibata, K., \& Ishii, T. 2010, \pasj, 62, 939
\bibitem[{Priest \& Forbes}(2002)]{prie02}                           Priest, E. R., \& Forbes, T. G. 2002, \aapr, 10, 313
\bibitem[{Scherrer}{~et~al.}(1995)]{sche95}                          Scherrer, P. H., Bogart, R. S., Bush, R. I., et al. 1995, \solphys, 162, 129
\bibitem[{Schmieder} {et~al.}(1991)]{Sch91}                          Schmieder, B., Fontenla, J., \& Tandberg-Hanssen, E. 1991, \aap, 252, 343
\bibitem[{Shen}{~et~al.}(2010)]{shen10}                              Shen, Y. D., Li, K. J., Yang, L. H., Yang, J. Y., \& Jiang, Y. C. 2010, AcASn, 51, 151
\bibitem[{Tandberg-Hanssen}(1995)]{tand95}                           Tandberg-Hanssen, E. 1995, The Nature of Solar Prominences, Dordrecht: Kluwer Acad. Publ
\bibitem[{Tang}(1987)]{tang87}                                       Tang, F. 1987, \solphys, 107, 233
\bibitem[{Thompson}{~et~al.}(2003)]{thom03}                          Thompson, W. T., Davila, J. M., Fisher, R. R., et al. 2003, \procspie, 4853, 1
\bibitem[{T$\ddot{\rm o}$r$\ddot{\rm o}$k \& Kliem}(2005)]{toro05}   T$\ddot{\rm o}$r$\ddot{\rm o}$k, T., \& Kliem, B. 2005, \apj, 630, L97
\bibitem[{Tripathi}{~et~al.}(2009)]{trip09}                          Tripathi, D., Gibson, S. E., Qiu, J., Fletcher, L., Liu, R., Gilbert, H., \& Mason, H. E. 2009, \aap, 498, 295
\bibitem[{Wang \& Liu}(2010)]{wang10}                                Wang, H. M., \& Liu, C. 2010, \apj, 716, L195
\bibitem[{Wheatland}{~et~al.}(2000)]{whea00}                         Wheatlang, M. S., Sturrock, P. A., \& Roumeliotis, G. 2000, \apj, 540, 1150
\bibitem[{W$\ddot{\rm u}$elser} {et~al.}(2004)]{Wue04}               W$\ddot{\rm u}$elser, J.-P, Lemen, J. R., Tarbell, T. D., et al. 2004, \procspie, 5171, 111
\bibitem[{Yang}{~et~al.}(2008)]{yang08}                              Yang, L. H., Jiang, Y. C., \& Ren, D. B. 2008, \chjaa, 8, 329
\bibitem[{Zhang}(2002)]{zhang02}                                     Zhang, H. Q. 2002, \mnras, 332, 500
\bibitem[{Zhang}{~et~al.}(2001)]{zhan01}                             Zhang, J., Dere, K. P., Howard, R. A., Kundu, M. R., \& White, S. M. 2001, 559, 452
\end{thebibliography}
\end{document}